\documentclass[epj]{webofc}
\usepackage[utf8]{inputenc}
\usepackage[varg]{txfonts}   % Web of Conferences font
\usepackage{booktabs}
\usepackage{xcolor}
\definecolor{darkred}{rgb}{0.4,0.0,0.0}
\definecolor{darkgreen}{rgb}{0.0,0.4,0.0}
\definecolor{darkblue}{rgb}{0.0,0.0,0.4}
\usepackage[bookmarks,linktocpage,colorlinks,
    linkcolor = darkred,
    urlcolor  = darkblue,
    citecolor = darkgreen]{hyperref}
\usepackage{subfigure}
\wocname{EPJ Web of Conferences}
\woctitle{Lattice2017}
\newcommand{\Slash}[1]{{\ooalign{\hfil/\hfil\crcr$#1$}}}

\newcommand{\eqn}[1]{(\ref{#1})}

\def\simge{\mathrel{%
       \rlap{\raise 0.511ex \hbox{$>$}}{\lower 0.511ex \hbox{$\sim$}}}}
\def\simle{\mathrel{
       \rlap{\raise 0.511ex \hbox{$<$}}{\lower 0.511ex \hbox{$\sim$}}}}

\begin{document}
%%%%%%%%%%%%%%%%%%%%%%%%%%%%%%%%%%%%%%%%%%%%%%%%%%%%%%%%%%%%%%%%%%%%%%%%%%%%%
%
\selectlanguage{english}
%----------------------------------------------------------------------------
\title{%
Equation of state in (2+1)-flavor QCD at physical point with improved Wilson fermion action using gradient flow
\footnote{Talk presented at the 35th International Symposium on Lattice Field Theory (LATTICE 2017), 18-24 June 2017, Granada, Spain.}
}
%----------------------------------------------------------------------------
\author{%
\firstname{Kazuyuki} \lastname{Kanaya}\inst{1}\fnsep\thanks{Speaker, \email{kanaya@ccs.tsukuba.ac.jp}, \\present address: Tomonaga Center for the History of the Universe, University of Tsukuba, Tsukuba, Ibaraki 305-8571, Japan}\and
\firstname{Shinji} \lastname{Ejiri}\inst{2}\and
\firstname{Ryo}  \lastname{Iwami}\inst{3}\and
\firstname{Masakiyo}  \lastname{Kitazawa}\inst{4,5}\and
\firstname{Hiroshi}  \lastname{Suzuki}\inst{6}\and
\firstname{Yusuke}  \lastname{Taniguchi}\inst{7}\and
\firstname{Takashi}  \lastname{Umeda}\inst{8}  \hspace{1mm} [WHOT-QCD Collaboration]
% etc.
}
%----------------------------------------------------------------------------
\institute{%
Center for Integrated Research in Fundamental Science and Engineering (CiRfSE), University of Tsukuba, Tsukuba, Ibaraki 305-8571, Japan
\and
Department of Physics, Niigata University, Niigata, Niigata 950-2181, Japan
\and
Track Maintenance of Shinkansen, Rail Maintenance 1st Department, East Japan Railway Company Niigata Branch, Niigata, Niigata 950-0086, Japan
\and
Department of Physics, Osaka University, Osaka, Osaka 560-0043, Japan
\and
J-PARC Branch, KEK Theory Center, Institute of Particle and Nuclear Studies, KEK, 203-1, Shirakata, Tokai, Ibaraki 319-1106, Japan
\and
Department of Physics, Kyushu University, 744 Motooka, Fukuoka, Fukuoka 819-0395, Japan
\and
Center for Computational Sciences (CCS), University of Tsukuba, Tsukuba, Ibaraki 305-8571, Japan
\and
Graduate School of Education, Hiroshima University, Higashihiroshima, Hiroshima 739-8524, Japan
}
%----------------------------------------------------------------------------
\abstract{%
We study the energy-momentum tensor and the equation of state as well as the chiral condensate in (2+1)-flavor QCD at the physical point applying the method of Makino and Suzuki based on the gradient flow. 
We adopt a nonperturbatively O($a$)-improved Wilson quark action and the renormalization group-improved Iwasaki gauge action. 
At Lattice 2016, we have presented our preliminary results of our study in (2+1)-flavor QCD at a heavy $u,d$ quark mass point. 
We now extend the study to the physical point and perform finite-temperature simulations in the range $T \simeq 155$--544 MeV ($N_t = 4$--14 including odd $N_t$'s) at $a \simeq 0.09$ fm.
% based on the fixed-scale approach using zero-temperature physical point configurations generated by the PACS-CS Collaboration. 
We show our final results of the heavy QCD study and present some preliminary results obtained at the physical point so far.
\\ \\
Preprint numbers: UTHEP-705, J-PARC-TH-0109, KYUSHU-HET-180, UTCCS-P-105
}

%----------------------------------------------------------------------------
\maketitle
%----------------------------------------------------------------------------
\section{Introduction}\label{intro}

The Yang-Mills gradient flow~\cite{Narayanan:2006rf,Luscher:2009eq,Luscher:2010iy,Luscher:2011bx,%
Luscher:2013cpa} has introduced big advances in lattice QCD. 
Fields at finite flow time, $t>0$, can be viewed as smeared
fields averaged over a physical radius of~$\sqrt{8t}$, and the operators constructed by flowed fields 
are shown to be free from UV divergences nor short-distance singularities. 
Because the flowed fields are defined nonperturbatively, we can treat the flowed operators as nonperturbatively renormalized ones whose finite expectation values can be calculated directly on the lattice.
This opened us a series of possibilities to significantly simplify the determination of physical observables on the lattice~\cite{Luscher:2013vga,Ramos:2015dla,Lat16suzuki}. 

In Ref.~\cite{Suzuki:2013gza},  a new method to calculate the energy-momentum tensor (EMT) on the lattice was proposed: 
To avoid dificulties due to explicit violation of the Poincare invariance on the lattice, EMT is defined in a continuum scheme by the Ward-Takahashi identities associated with the Poincare transformation. 
When we flow the system, the flowed EMT operator becomes calculable on the lattice, but some unwanted operators can contaminate at $t>0$ --- as a consequence, e.g., the flowed EMT does not satisfy the WT identities.
However, such unwanted contributions % are absent in the $t\to0$ limit and thus 
can be removed by taking a $t\to0$ extrapolation.
The extrapolation can be made smoother by using a small-$t$ operator expansion, because the mixing coefficients can be calculated by perturbation theory in asymptotically free theories~\cite{Makino:2014taa,Suzuki:2013gza}.

The equation of state (EOS) can be extracted from the diagonal components of the EMT. 
We note that this method does not require the information of beta functions, which is sometimes a big burden in conventional evaluation of EOS on the lattice using the derivative or ($T$-)integration methods.

The new method has been shown to be powerful in quenched QCD by the FlowQCD Collaboration \cite{Asakawa:2013laa}.  
We are extending the study to (2+1)-flavor QCD adopting the Iwasaki gauge action~\cite{Iwasaki:2011np} and a non-perturbatively $O(a)$-improved Wilson quark action~\cite{Sheikholeslami:1985ij}.
As the first step, we studied the case of heavy $u$ and $d$ quarks with approximately physical $s$ quark.
Some preliminary results were presented at the last lattice conference~\cite{Lat16WHOTa,Lat16WHOTb}.
After the conference, we have made a series of additional tests to confirm the procedure, and started a study of (2+1)-flavor QCD just at the physical point.
In this paper, we show the finial results of the heavy QCD study~\cite{WHOT2017b,WHOT2017}, and some preliminary results obtained at the physical point.
See \cite{taniguchi2017} for another development of the study.
%further application of the method in (2+1)-flavor QCD.

%----------------------------------------------------------------------------
\section{Energy-momentum tensor on the lattice}\label{sec:emt}

The gradient flow we adopt is the simplest one: 
The gauge flow is given in \cite{Luscher:2010iy} and
the quark flow is given in \cite{Luscher:2013cpa}.
Besides wave function renormalization of the quark fields, this flow preserves the finiteness of the flowed operators \cite{Luscher:2013cpa}.

According to Refs.~\cite{Suzuki:2013gza,Makino:2014taa}, the correctly normalized EMT is given by 
\begin{align}
   T_{\mu\nu}(x)
   &=\lim_{t\to0}\biggl\{c_1(t)\left[
   \Tilde{\mathcal{O}}_{1\mu\nu}(t,x)
   -\frac{1}{4}\Tilde{\mathcal{O}}_{2\mu\nu}(t,x)
   \right]
%\notag\\
%   &\qquad{}
   +c_2(t)\left[
   \Tilde{\mathcal{O}}_{2\mu\nu}(t,x)
   -\left\langle\Tilde{\mathcal{O}}_{2\mu\nu}(t,x)\right\rangle_{\! 0}
   \right]
\notag\\
   &%\qquad{}
   +c_3(t)\sum_{f=u,d,s}
   \left[
   \Tilde{\mathcal{O}}_{3\mu\nu}^f(t,x)
   -2\Tilde{\mathcal{O}}_{4\mu\nu}^f(t,x)
   -\left\langle
   \Tilde{\mathcal{O}}_{3\mu\nu}^f(t,x)
   -2\Tilde{\mathcal{O}}_{4\mu\nu}^f(t,x)
   \right\rangle_{\! 0}
   \right]
\label{eq:(2.8)}\\
   &
   +c_4(t)\sum_{f=u,d,s}
   \left[
   \Tilde{\mathcal{O}}_{4\mu\nu}^f(t,x)
   -\left\langle\Tilde{\mathcal{O}}_{4\mu\nu}^f(t,x)\right\rangle_{\! 0}
   \right]
   +\sum_{f=u,d,s}c_5^f(t)\left[
   \Tilde{\mathcal{O}}_{5\mu\nu}^f(t,x)
   -\left\langle\Tilde{\mathcal{O}}_{5\mu\nu}^f(t,x)\right\rangle_{\! 0}
   \right]\biggr\},
\notag
\end{align}
with $\langle\cdots\rangle_{0}$ standing for the expectation value at $T=0$ 
and
$   \Tilde{\mathcal{O}}_{1\mu\nu}\equiv
   G_{\mu\rho}^a\,G_{\nu\rho}^a$,
$   \Tilde{\mathcal{O}}_{2\mu\nu}\equiv
   \delta_{\mu\nu}\,G_{\rho\sigma}^a\,G_{\rho\sigma}^a$,
$   \Tilde{\mathcal{O}}_{3\mu\nu}^f\equiv
   \varphi_f(t)\,\Bar{\chi}_f
   \left(\gamma_\mu\overleftrightarrow{D}_\nu
   +\gamma_\nu\overleftrightarrow{D}_\mu\right)
   \chi_f$,
$   \Tilde{\mathcal{O}}_{4\mu\nu}^f\equiv
   \varphi_f(t)\,\delta_{\mu\nu}\,
   \Bar{\chi}_f
   \overleftrightarrow{\Slash{D}}
   \chi_f$,
and
$   \Tilde{\mathcal{O}}_{5\mu\nu}^f\equiv
   \varphi_f(t)\,\delta_{\mu\nu}\,
   \Bar{\chi}_f\,
   \chi_f$,
where
$
   \overleftrightarrow{D}_\mu\equiv D_\mu-\overleftarrow{D}_\mu
$
and the quark normalization factor~$\varphi_f(t)$ is given by~\cite{Makino:2014taa}
\begin{equation}
   \varphi_f(t)\equiv
  - 6
   \, \left[(4\pi)^2\,t^2
   \left\langle\Bar{\chi}_f(t,x)\overleftrightarrow{\Slash{D}}\chi_f(t,x)
   \right\rangle_{\! 0}\right]^{-1} .
\label{eq:(2.15)}
\end{equation}
%We then have $p/T^4  = \sum_i \langle T_{ii}\rangle /(3T^4)$, $\epsilon/T^4 = - \langle T_{00}\rangle/T^4$ for EOS.
The coefficients $c_i(t)$ %in Eq.~\eqref{eq:(2.8)} 
are calculated by perturbation theory in~\cite{Makino:2014taa}.
%\begin{align}
%   c_1(t)&
%   =\frac{1}{\Bar{g}\!\left(1/\sqrt{8t}\right)^2}
%   -\frac{1}{(4\pi)^2}\left[
%   9(\gamma-2\ln2)+\frac{19}{4}\right],
%\qquad{}
%   c_2(t)
%  =\frac{1}{(4\pi)^2}\frac{33}{16},
%\notag 
%\\
%   c_3(t)&
%   =\frac{1}{4}\left\{1+\frac{\Bar{g}\!\left(1/\sqrt{8t}\right)^2}{(4\pi)^2}
%   \left[2+\frac{4}{3}\ln(432)\right]\right\},
%\qquad{}
%   c_4(t)=\frac{1}{(4\pi)^2}\Bar{g}\!\left(1/\sqrt{8t}\right)^2,
%\notag 
%\\
%   c_5^f(t)&=-\Bar{m}_f\!\left(1/\sqrt{8t}\right)
%   \left\{1+\frac{\Bar{g}\!\left(1/\sqrt{8t}\right)^2}{(4\pi)^2}
%   \left[4(\gamma-2\ln2)+\frac{14}{3}+\frac{4}{3}\ln(432)\right]\right\},
%\notag 
%\end{align}
%where $\gamma$ is the Euler constant and $\Bar{g}(\mu)$
%and~$\Bar{m}_f(\mu)$ are the running gauge coupling and the running quark mass
%of the flavor~$f$ in the $\overline{\mathrm{MS}}$ scheme at the scale $\mu$.
%Though these coefficients are calculated by perturbation theory, they are used just to guide the $t\to0$ extrapolation. We thus consider that our evaluation is essentially non-perturbative. 
%
%We also emphasize that the non-perturbative beta functions or Karsch coefficients, which require a big numerical task in conventional calculations of EOS in particular in full QCD, are not needed. % in the gradient flow method. %This is a great benefit of the new method.

As discussed in the introduction, we need to carry out the extrapolation $t\to0$ to remove contamination of unwanted operators.
In \eqn{eq:(2.8)}, the continuum extrapolation $a\to0$ is assumed to have been done before.
In numerical studies, however, it is often favorable to take the continuum extrapolation at a later stage of analyses.
On finite lattices with $a\ne0$, we expect additional contamination of unwanted operators.
Since we adopt the non-perturbatively $O(a)$-improved Wilson quarks,
the lattice artifacts start with $O(a^2)$
and we expect
\begin{equation}
   T_{\mu\nu}(t,x,a)
   =T_{\mu\nu}(x) + t S_{\mu\nu}(x)
   +A_{\mu\nu}\frac{a^2}{t}+\sum_f B^f_{\mu\nu}(am_f)^2+C_{\mu\nu}(aT)^2
   +D_{\mu\nu}(a\Lambda_{\mathrm{QCD}})^2
   +a^2S'_{\mu\nu}(x)+O(a^4,t^2)
\label{eq:a2overt}
\end{equation}
where $T_{\mu\nu}(x)$ is the physical EMT, $S_{\mu\nu}$ and $S'_{\mu\nu}$ are contaminations of dimension-six operators with the same quantum number, and 
$A_{\mu\nu}$, $B^f_{\mu\nu}$, $C_{\mu\nu}$, and~$D_{\mu\nu}$ are those from dimension-four operators. 
To exchange the order of the limiting procedures $a\to0$ and $t\to0$, the singular terms like $a^2/t$ must be removed. 
This is possible if we have a window in $t$ in which the linear terms dominate.

%----------------------------------------------------------------------------
\section{QCD with heavy ud quarks}\label{sec:heavy}

As the first application of the method to full QCD, we study the case with heavy $u$ and $d$ quarks.
The zero temperature configurations were generated on a $28^3\times56$ lattice 
at $a=0.0701(29)\,\mathrm{fm}$ ($1/a\simeq2.79\,\mathrm{GeV}$) 
with heavy $u$ and $d$ quarks, $m_\pi/m_\rho\simeq0.63$, and
almost physical $s$ quark, $m_{\eta_{ss}}/m_\phi\simeq0.74$ \cite{Ishikawa:2007nn}.
Adopting the fixed-scale approach~\cite{Levkova:2006gn,Umeda:2008bd},
corresponding finite-temperature configurations were generated
on $32^3\times N_t$ lattices with $N_t=16$, 14, $\cdots$ 4 ($T\approx174$--697 MeV) 
for a study of EOS using the conventional $T$-integration method~\cite{Umeda:2012er}. 
The pseudocritical temperature was estimated to be $T_\textrm{pc} \sim 190$ MeV \cite{Umeda:2012er}.

\begin{figure}[htb]
\centering
\includegraphics[width=6cm,clip]{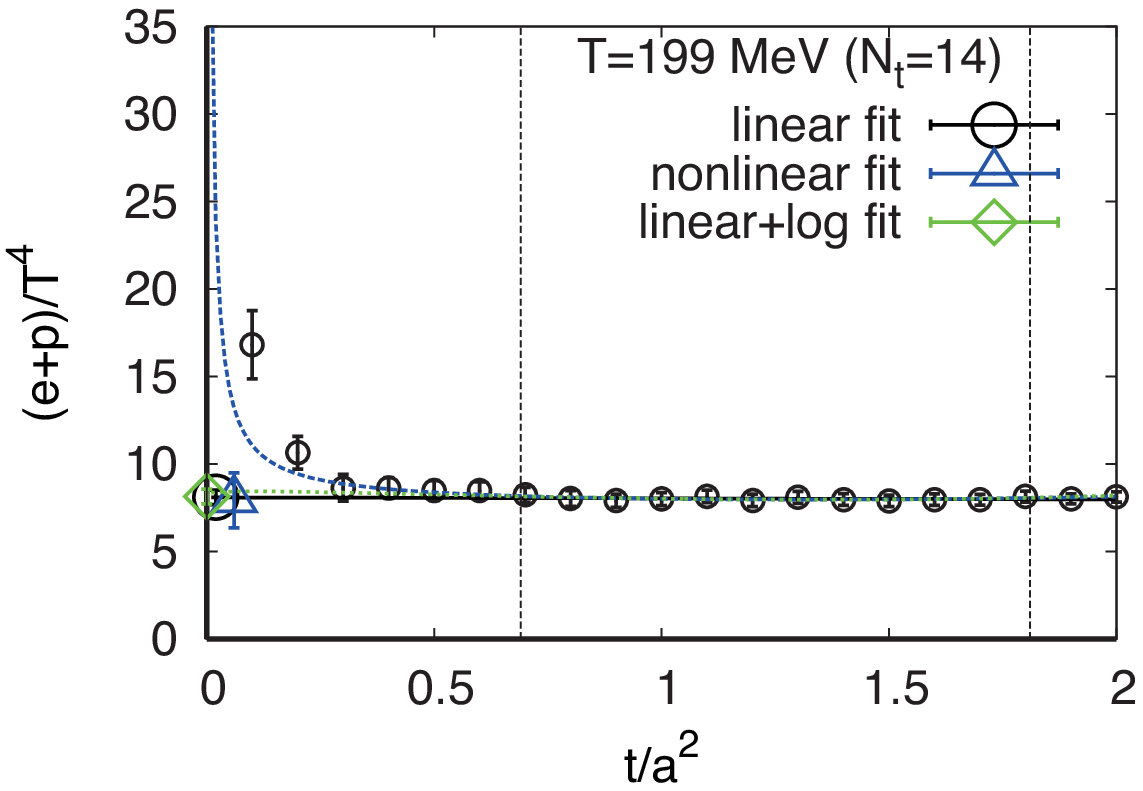}
\hspace{10mm}
\includegraphics[width=6cm,clip]{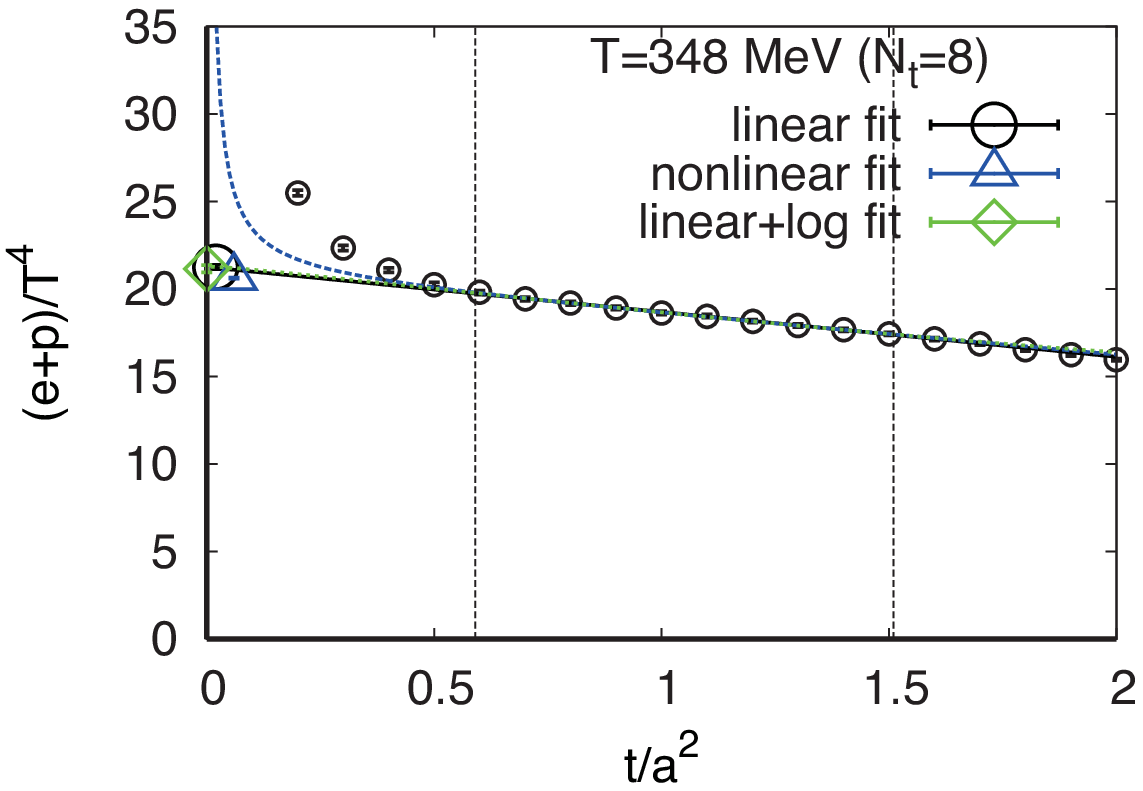}
\caption{Entropy density $(\epsilon+p)/T^4$ in heavy QCD as a function of the flow time. 
The pair of dashed vertical lines indicates the window used for the fit.
Black solid lines are the results of the linear fits, and the big open circles at~$t=0$ are the results of physical EOS.
Blue and green dashed curves together with blue upward triangles and green diamonds at $t\sim0$ are the fit results with the nonlinear ansatz~\eqref{eqn:5.4} and linear+log ansatz~\eqref{eqn:5.4l}, respectively. 
Errors are statistical only. 
See \cite{WHOT2017b} for other temperatures.
}
\label{fig5}
\end{figure}

\begin{figure}[htb]
\centering
\includegraphics[width=6cm,clip]{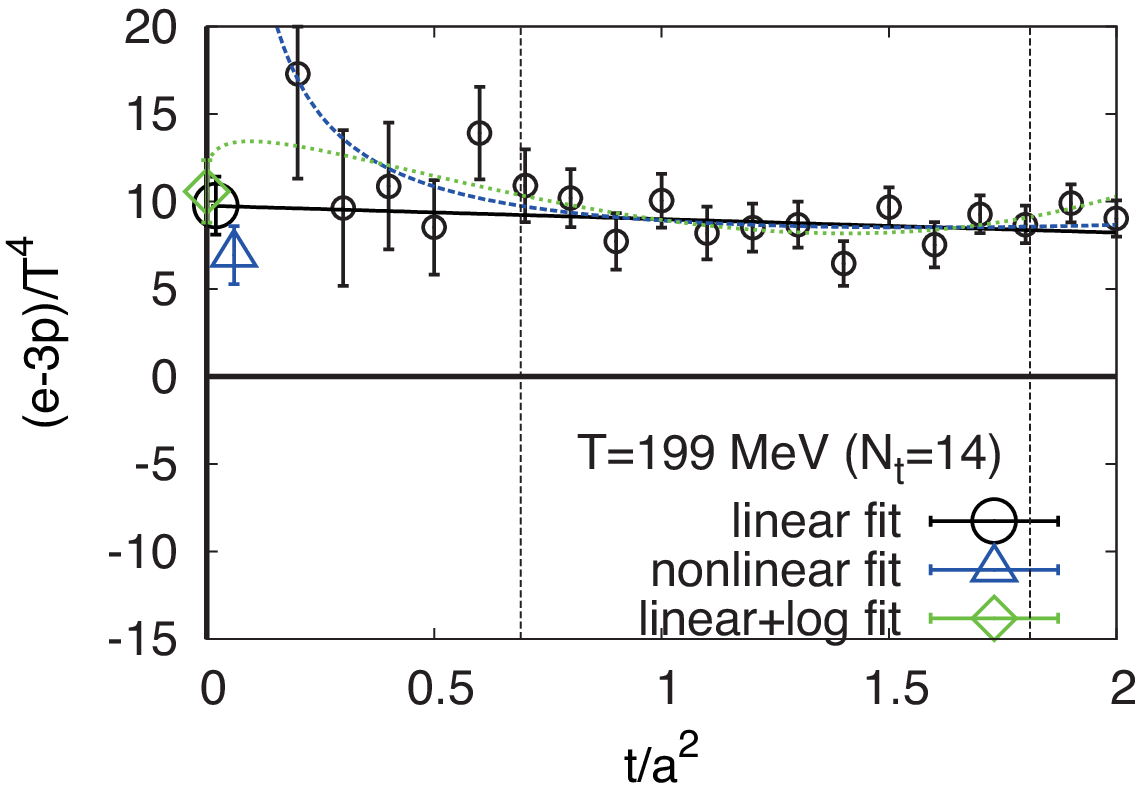}
\hspace{8mm}
\includegraphics[width=6cm,clip]{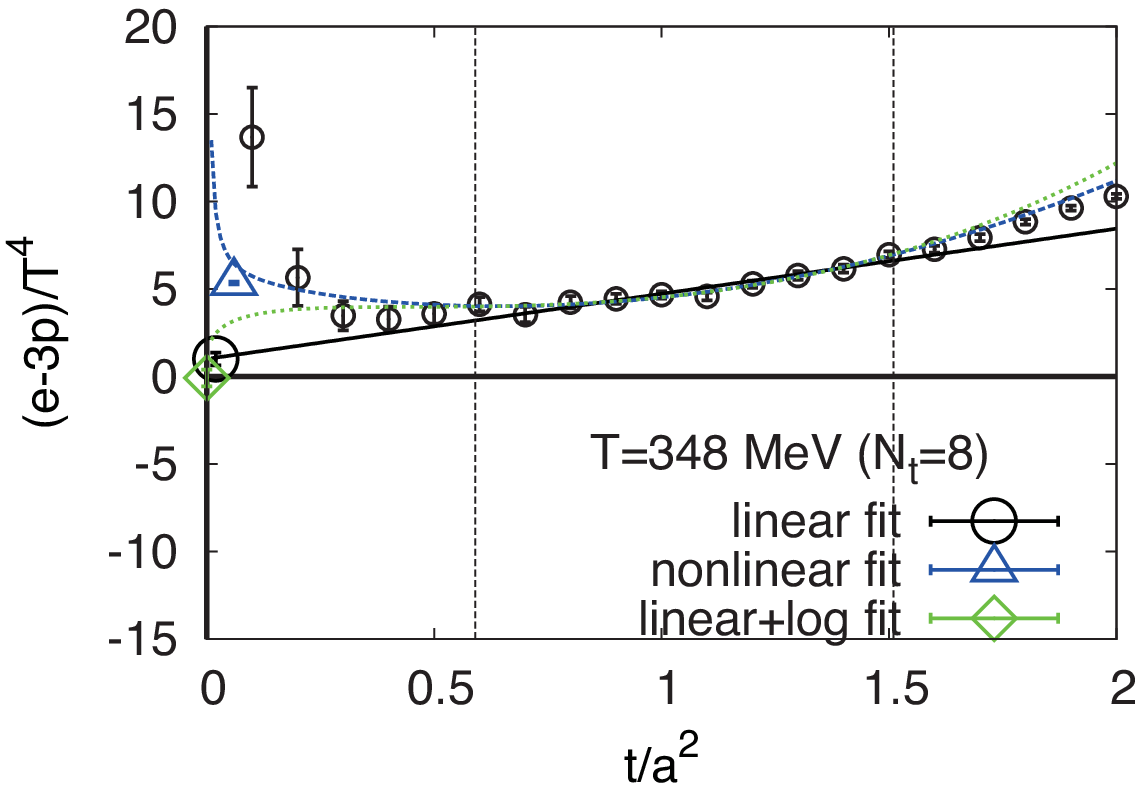}
\caption{The same as Fig.~\ref{fig5} but for the trace anomaly~$(\epsilon-3p)/T^4$ in heavy QCD.
See \cite{WHOT2017b} for other temperatures.
}
\label{fig5b}
\end{figure}

Preliminary results of the study were presented at the previous lattice conference~\cite{Lat16WHOTa,Lat16WHOTb}.
We found that, though the EMT data show the $a^2/t$ singularity at small $t$, the data show wide windows in $t$ in which the EMT is well linear in $t$.
We have thus decided to extrapolate the data to $t=0$ using data within the window
except for the case of $T\simeq697$ MeV ($N_t=4$) for which a clear linear window could not be identified.

After the previous conference, we have carried out a series of additional analyses to check the validity of the linear extrapolation procedure.
In particular, in addition to (i) the original linear fit,  we have done fits adopting 
(ii) nonliner ansatz
\begin{equation}
 \left\langle T_{\mu\nu}(t,a) \right\rangle
  =\langle T_{\mu\nu} \rangle +A_{\mu\nu}\frac{a^2}{t}+t\,S_{\mu\nu}
  + t^2 R_{\mu\nu} 
%  + O(a^2,t^3)
\label{eqn:5.4}
\end{equation}
inspired from higher order terms in $t$ and $a$ in \eqn{eq:a2overt}, and (iii) linear+log ansatz
\begin{equation}
 \left\langle T_{\mu\nu}(t,a) \right\rangle
  =\langle T_{\mu\nu} \rangle + t\,S_{\mu\nu}
  +   \left[ \log(\sqrt{8t}/a) \right]^{-2} Q_{\mu\nu}
\label{eqn:5.4l}
\end{equation}
inspired from higher order terms in the coefficients $c_i$.
A fit including all the correction terms in (\ref{eqn:5.4}) and (\ref{eqn:5.4l}) turned out to be unstable due to too many fitting parameters.
Typical results of the nonlinear and liner+log fits for the entropy density and the trace anomaly using the same linear window are shown by blue and green dashed curves in Figs.~\ref{fig5} and \ref{fig5b}. 
% As discussed in~\cite{Lat16WHOTa,WHOT2017b}, 
We find that, in most cases, the three fits lead to a consistent result.
We use the linear fit for the central value and take the differences among the fits as an estimate of the systematic error due to the extrapolation.

\begin{figure}[tb]
\centering
\includegraphics[width=6cm,clip]{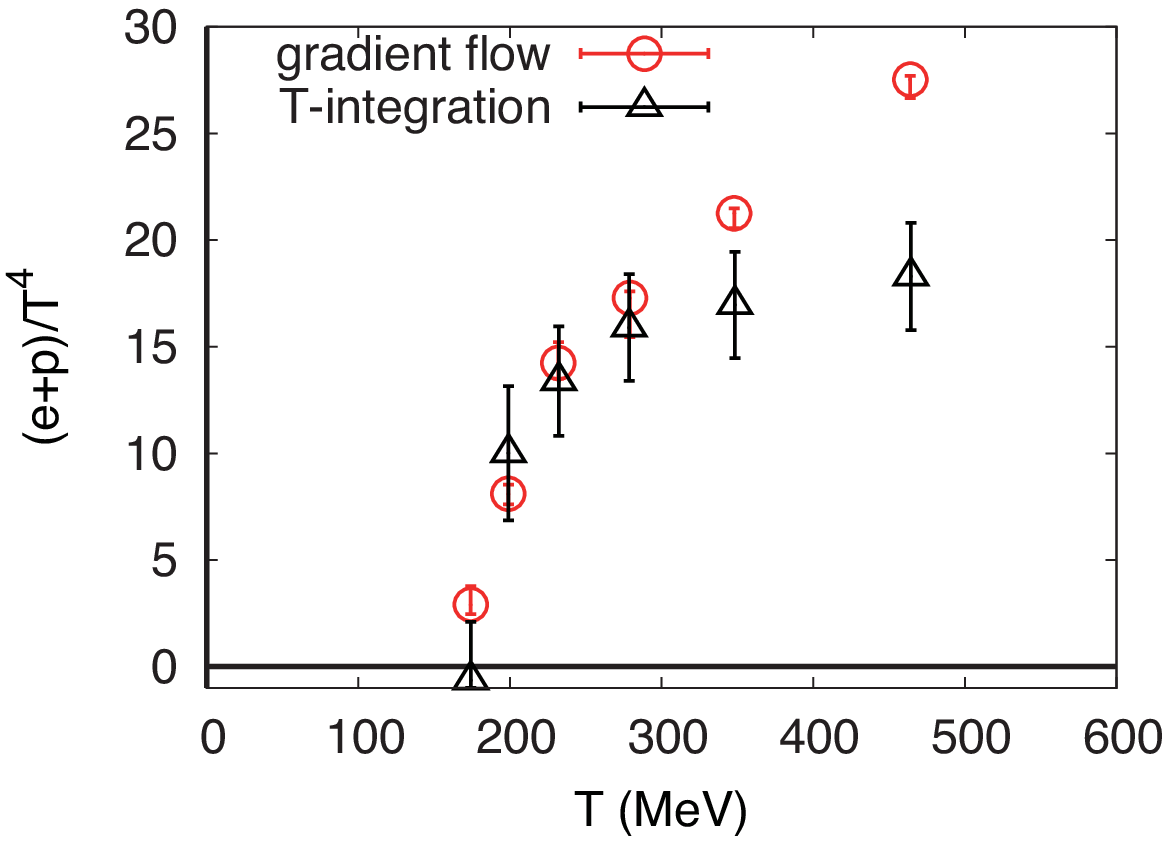}
\hspace{8mm}
\includegraphics[width=6cm,clip]{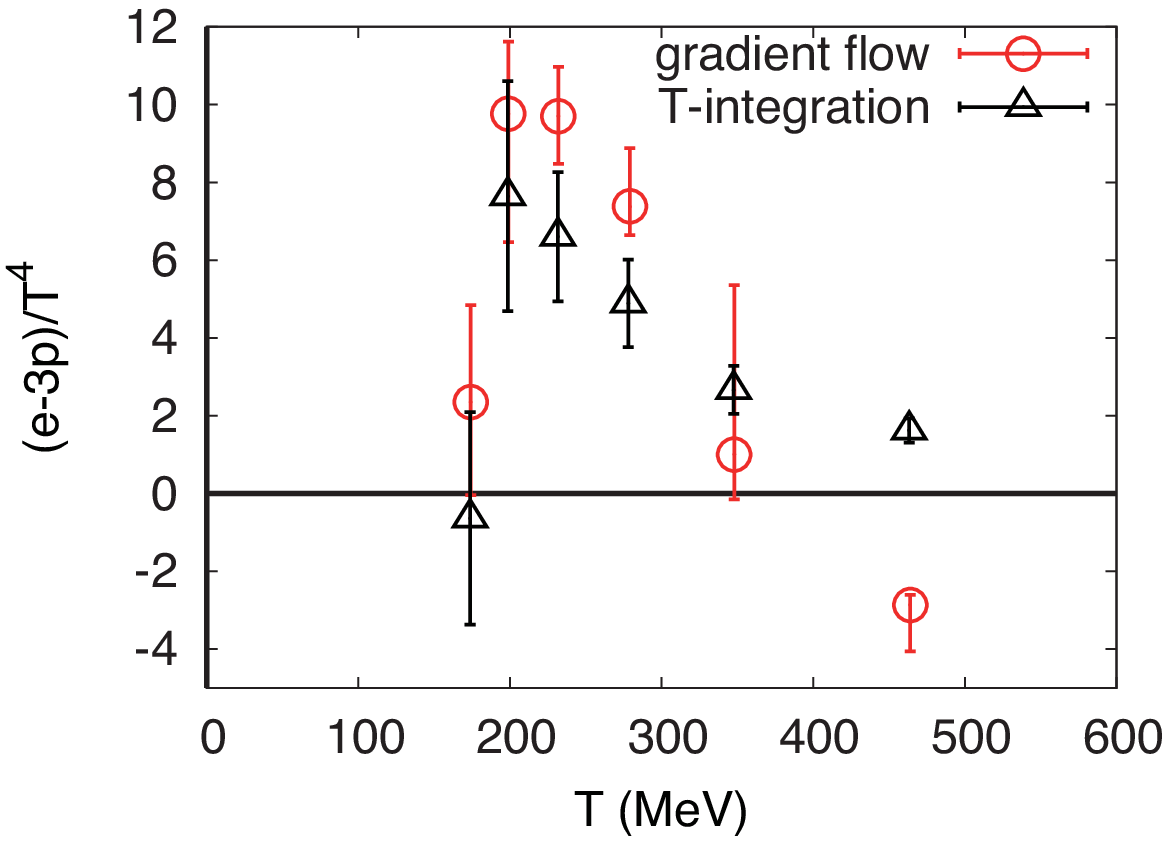}
\caption{Entropy density $(\epsilon+p)/T^4$ and  trace anomaly~$(\epsilon-3p)/T^4$ in heavy QCD 
as functions of temperature. 
Errors include both statistical and systematic errors. 
See \cite{WHOT2017b} for details.
}
\label{fig6}
\end{figure}

Our final results of EOS, 
with the energy density and the pressure determined by $\epsilon=-\langle T_{00}\rangle$ and $p=\sum_i \langle T_{ii}\rangle/3$,
are shown in Fig.~\ref{fig6}. 
Red circles are our result with the gradient flow method.  
Black triangles are previous results obtained by the $T$-integration method~\cite{Umeda:2012er}.
We find that the results of the gradient flow method are consistent
with those of the $T$-integration method at low
temperatures $T\simle280\,\mathrm{MeV}$ ($N_t \simge 10$). 
The deviations at $T\simge350\,\mathrm{MeV}$ ($N_t \simle 8$) may be due to a lattice
artifact of $O\left((aT)^2\right)=O\left(1/N_t^2\right)$ at small $N_t$

\begin{figure}[tb]
\centering
\includegraphics[width=6cm,clip]{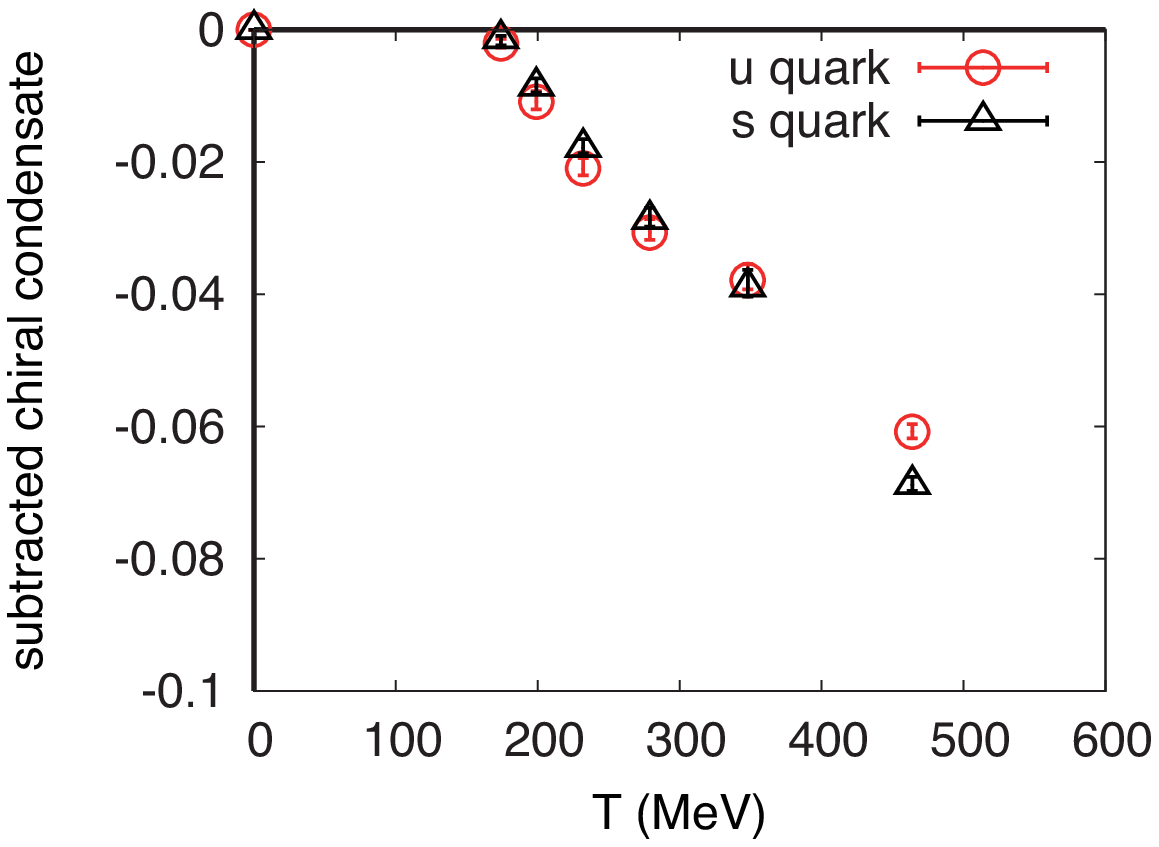}
\hspace{8mm}
\includegraphics[width=6cm,clip]{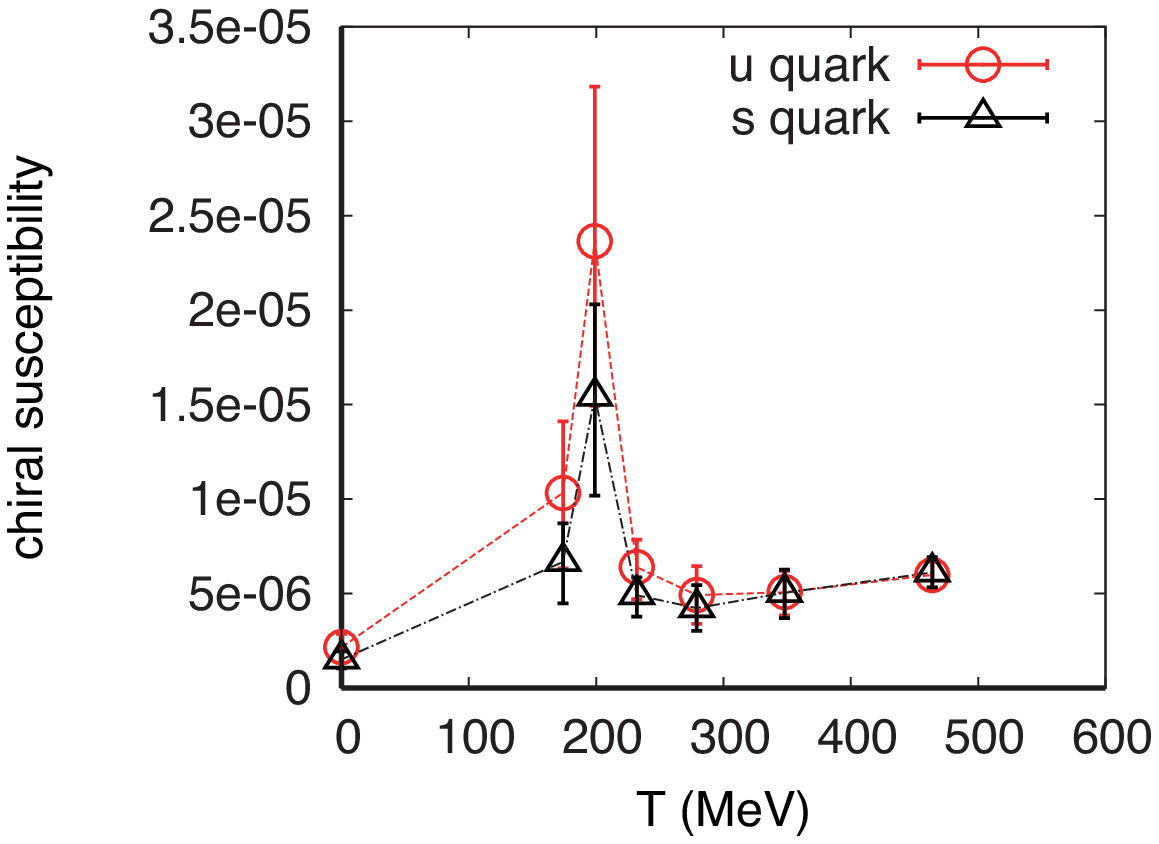}
\caption{
Renormalized chiral condensate with the VEV subtraction
$-\left\langle\{\Bar{\psi}_f\psi_f\}\right\rangle_{\overline{\mathrm{MS}}}$ 
and disconnected chiral susceptibility~$\chi_{\Bar{f}f}^{\mathrm{disc.}}$ in heavy QCD,
both in $\overline{\mathrm{MS}}$ scheme at $\mu\!=\!2\,\mathrm{GeV}$, as functions of temperature. 
The vertical axis is in unit of ${\rm GeV}^3$ and $\mathrm{GeV}^6$, respectively.
% Red circles are $u$ (or $d$) quark condensate and black triangles are that for $s$ quark. 
Errors include both statistical and systematic errors. 
See \cite{WHOT2017b} for details.
}
\label{fig_chiral}
\end{figure}

The method can be extended to other quark bilinear observables \cite{Hieda:2016lly}. 
Our results for the chiral condensate and its disconnected susceptibility are shown in Fig.~\ref{fig_chiral}. 
We see a clear signal of chiral crossover at $T \sim 190$ MeV, which is consistent with $T_\textrm{pc}$ suggested by Polyakov loop etc. \cite{Umeda:2012er}.
We note that dependence on the valence quark mass is small in the condensate.
On the other hand, the disconnected chiral susceptibility seems to show a higher peak as the quark mass is decreased.
See \cite{WHOT2017} for our results on the topological charge and its susceptibility. 
From these studies, we find that the gradient flow method is quite powerful in extracting physical properties even with Wilson-type quarks which violate the chiral symmetry explicitly and thus had not been easy to calculate chiral and topological quantities with usual methods.

%----------------------------------------------------------------------------
\section{QCD at the physical point}\label{sec:details}

\begin{figure}[tb]
\centering
\includegraphics[width=4cm,clip]{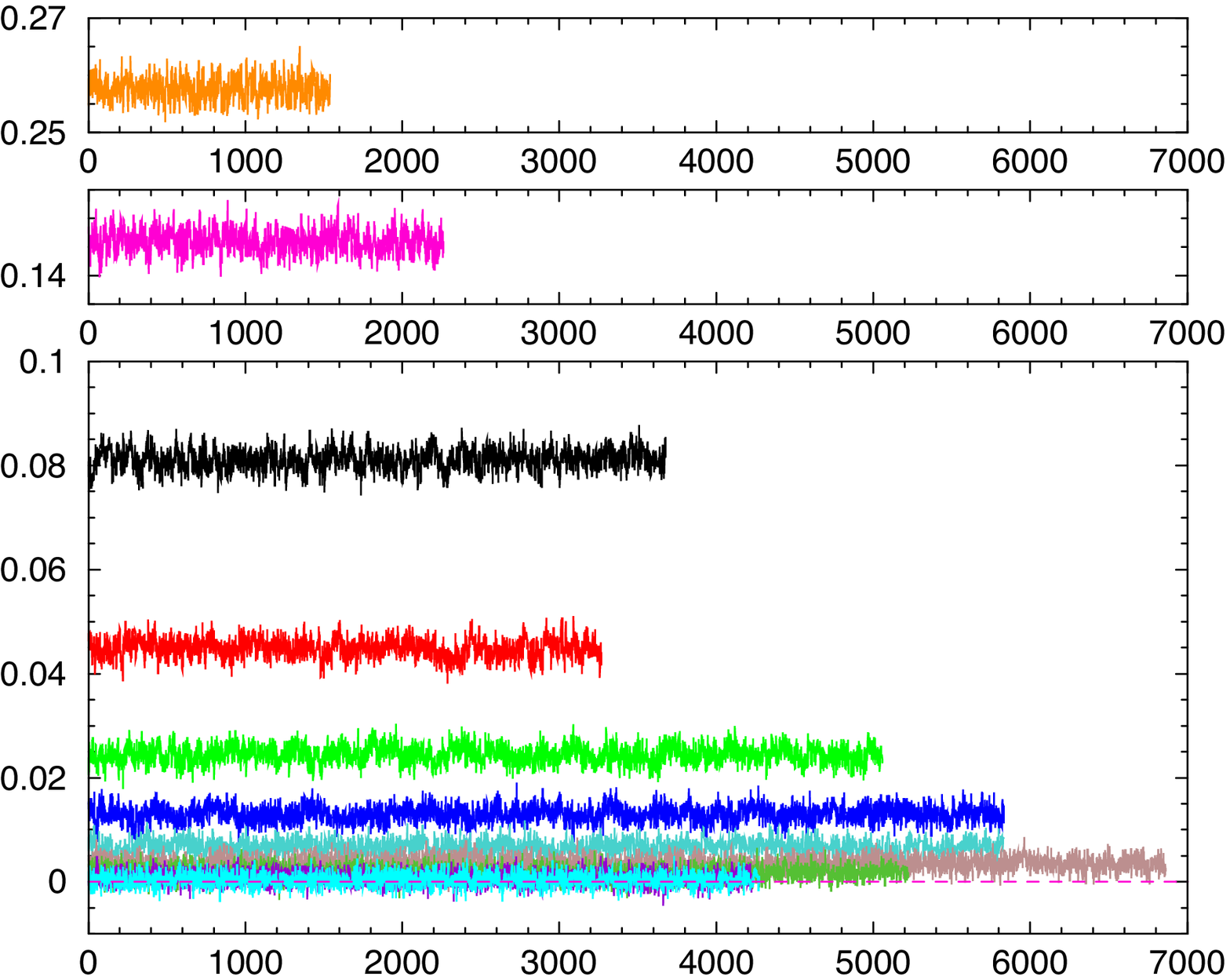}
\hspace{1mm}
\includegraphics[width=4.6cm,clip]{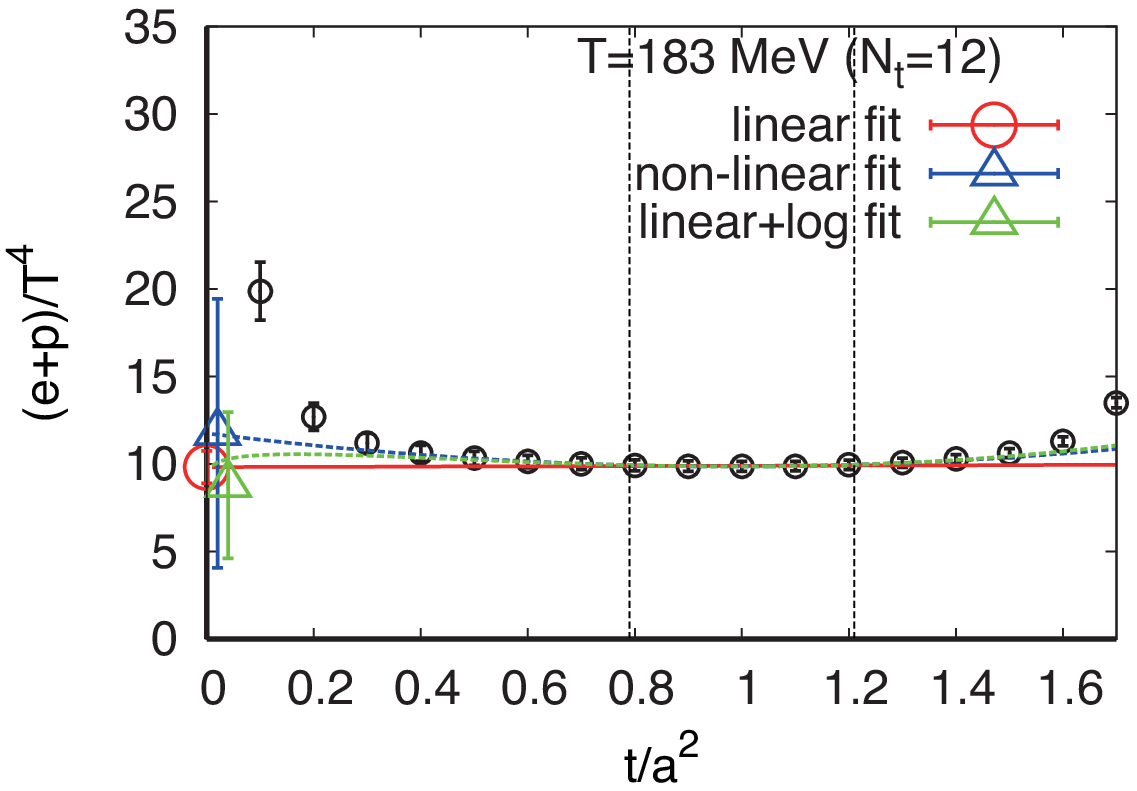}
\hspace{1mm}
\includegraphics[width=4.6cm,clip]{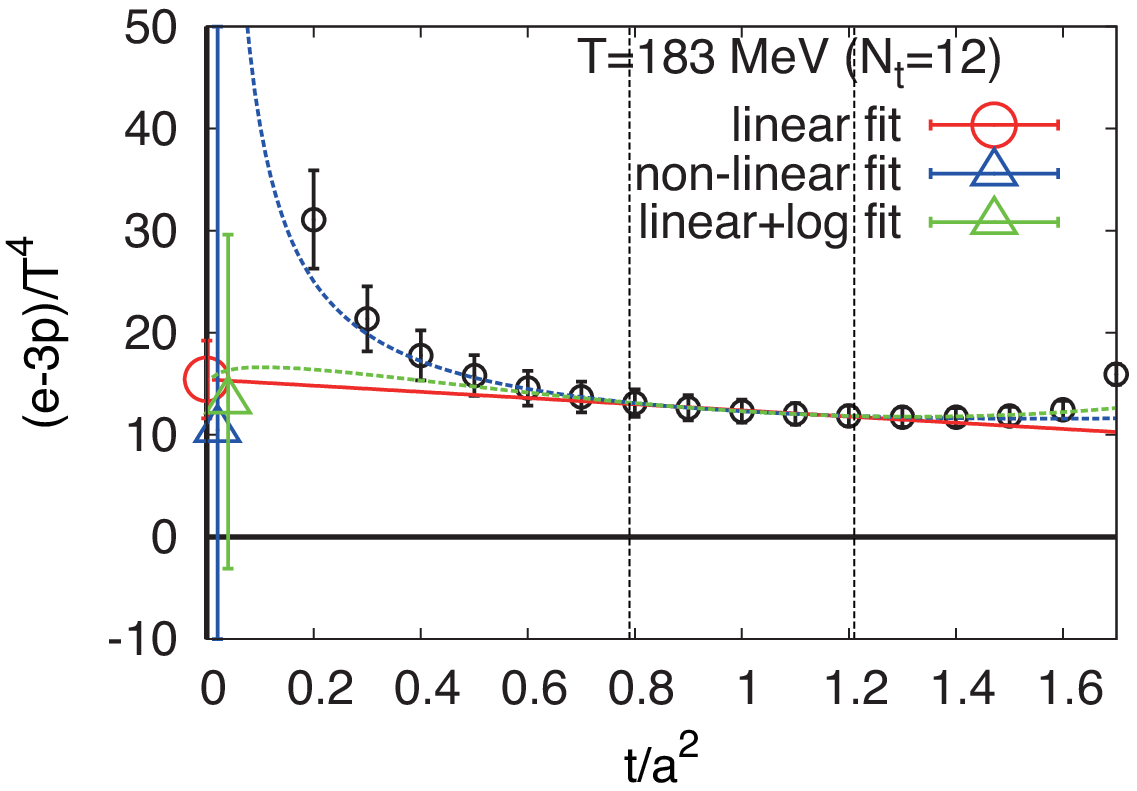}
\caption{
\textbf{\em Left}: 
Time history of the Polyakov loop obtained on our finite temperature lattices at the physical point. 
In the top and middle panels, the histories of $N_t=4$ and 5 simulations are shown, respectively.
In the panel at the bottom, the histories at $N_t=6$, 7, $\cdots$, 14 are shown from top to bottom. 
The horizontal axis is the trajectory length in MD time units.
\textbf{\em Middle}: 
Preliminary fit results of the entropy density as functions of the flow time on $N_t=12$ lattice at the physical point.
\textbf{\em Right}: 
The same as the middle panel but for the trace anomaly.
}
\label{fig_physpt}
\end{figure}

As one of the next steps, we are extending the study to $N_f=2+1$ QCD just at the physical point
using zero-temperature configurations generated by the PACS-CS Collaboration with Iwasaki gauge action and non-perturbatively improved Wilson quark action at $\beta=1.90$ on a $32^3\times64$ lattice \cite{Aoki:2009ix}.
The quark masses are fine-tuned to the physical point by a reweighting technique. The lattice spacing is estimated to be $a=0.08995(40)$ fm, and the spatial lattice size of $N_s=32$ corresponds to 3 fm.
We are generating finite-temperature gauge configurations directly at the physical point 
on $32^3 \times N_t$ lattices with $N_t=14$, 13, $\cdots$, 5, 4, which corresponds to the temperature range of $T \approx 160$--550 MeV \cite{umeda:lat15}.
Here, odd vales of $N_t$ are also simulated to achieve a finer resolution in $T$.
We note that the lattices are slightly coarser than the case of heavy QCD. 
Although $T_\textrm{pc}$ is not known yet with this lattice action at the physical point, we expect it to be smaller than the 190 MeV for the heavy QCD case.
Latest status of the simulations is shown in the left panel of Fig.~\ref{fig_physpt}.

\begin{figure}[tb]
\centering
\includegraphics[width=6cm,clip]{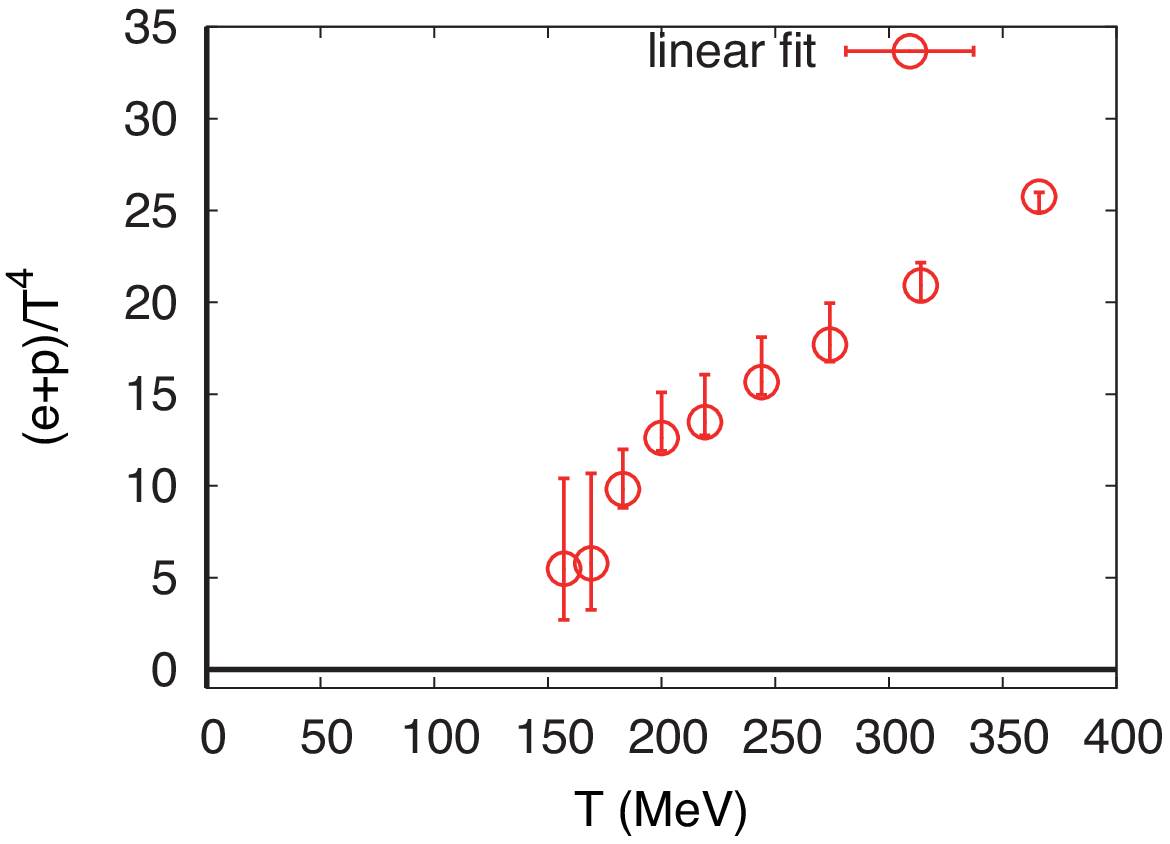}
\hspace{8mm}
\includegraphics[width=6cm,clip]{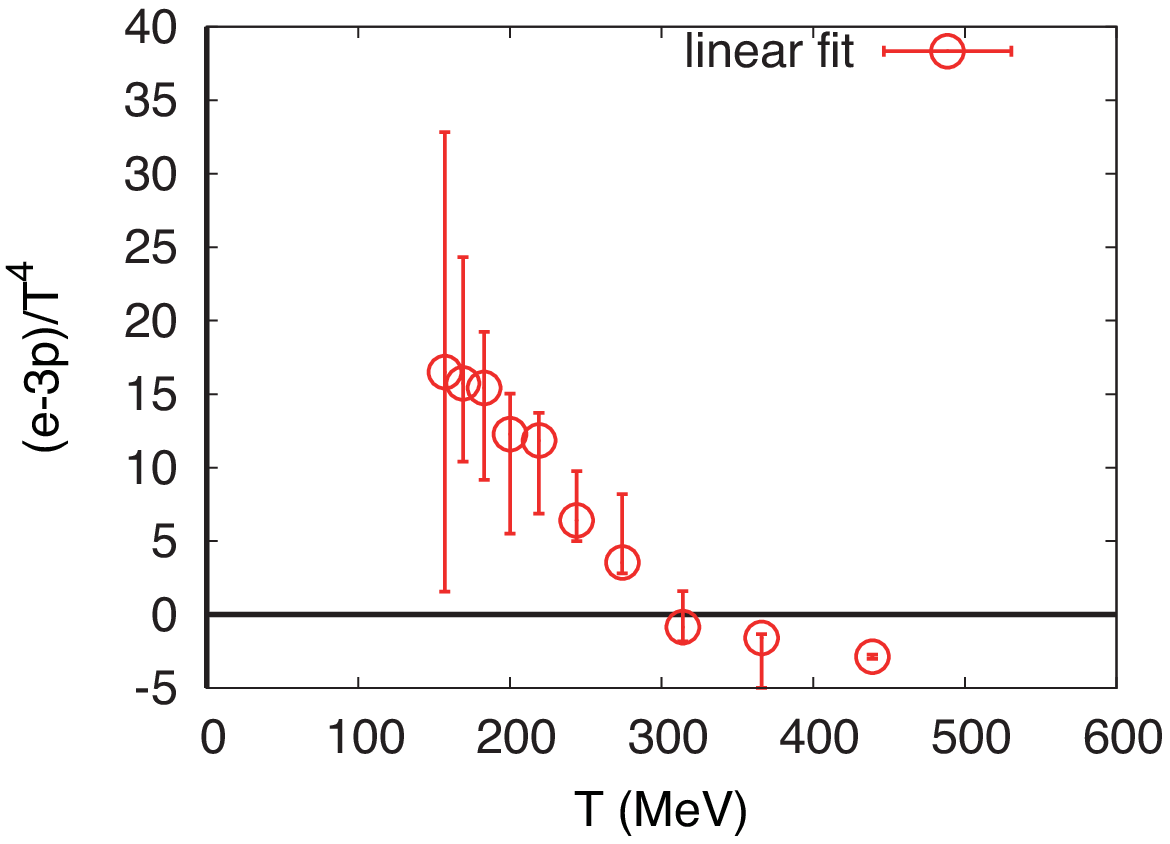}
%
%\vspace{-2mm}
\caption{
Preliminary results for the entropy density and  trace anomaly in physical point QCD 
as functions of temperature. 
Errors include both statistical and systematic errors. 
Data at $T\simge247$ MeV ($N_t\le8$) would be contaminated by the $O\left( (aT)^2\right)=O\left(1/N_t^2 \right)$ lattice artifacts.
}
\label{fig6_P}
\end{figure}

In the middle and right panels of Fig.~\ref{fig_physpt}, we show our preliminary results of the entropy density and the trace anomaly as functions of the flow time, computed using the configurations accumulated so far on the $N_t=12$ lattice. 
Results at other $N_t$'s are similar. 
We note that the linear windows at $N_t \ge 6$ are often narrower than the case of heavy QCD shown in Fig.~\ref{fig5}.
This may be caused by the coarser lattice spacing at the physical point and/or the lighter $u,d$ quark mass. 

In this trial study, we adopt the same strategy as in the heavy QCD: 
When we find a linear window containing five or more points within the statistical errors, we extrapolate the data to the $t\to0$ limit by the linear fit and take the differences with nonlinear and linear+log fits as an estimate of the systematic error due to the extrapolation.

Our preliminary results of EOS extrapolated to $t=0$ are shown in Fig.~\ref{fig6_P}.
From our experience in the heavy QCD, data at $T\simge247$ MeV would be contaminated by the $O\left( (aT)^2\right)=O\left(1/N_t^2 \right)$ lattice artifacts  on $N_t\le8$ lattices.
We find that the entropy density is well consistent with the results of staggered quarks at the physical point in the continuum limit \cite{Borsanyi:2013bia,Bazavov:2014pvz}, while the central values for the interaction measure are several times larger than those of the staggered quarks though our errors are still large.

\begin{figure}[tb]
\centering
\includegraphics[width=6cm,clip]{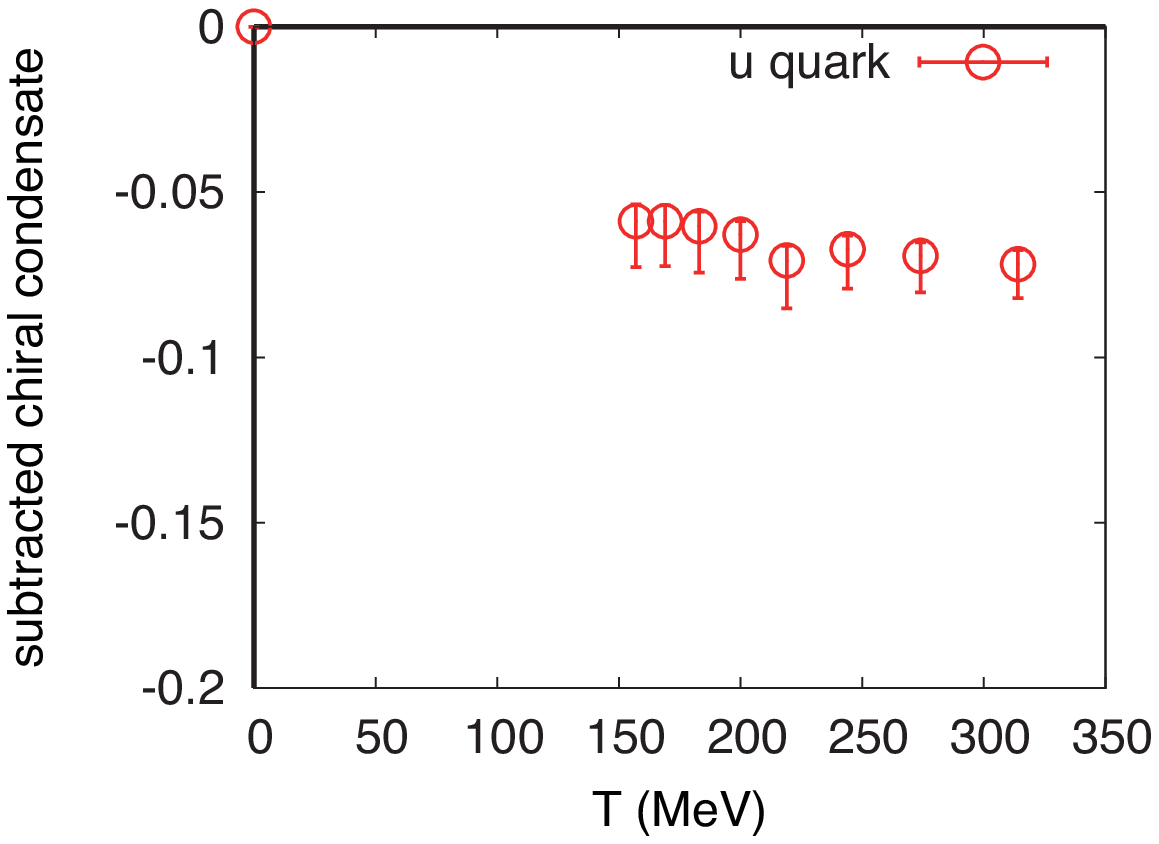}
\hspace{8mm}
\includegraphics[width=6cm,clip]{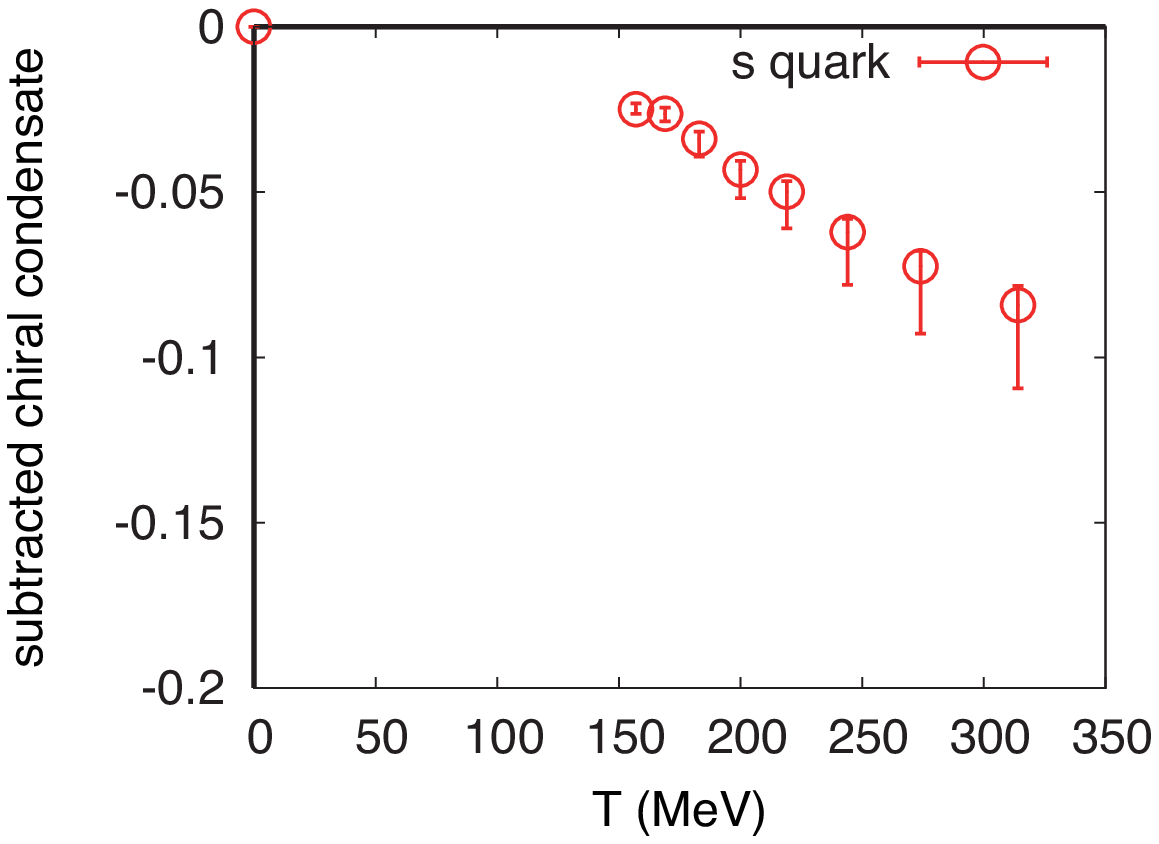}
%
%\vspace{-2mm}
\caption{
%Preliminary results 
The same as Fig.~\ref{fig6_P} but for the renormalized chiral condensate with the VEV subtraction
%$-\left\langle\{\Bar{\psi}_f\psi_f\}\right\rangle_{\overline{\mathrm{MS}}}$ 
for light quark (left) and $s$ quark (right) in $\overline{\mathrm{MS}}$ scheme at $\mu\!=\!2\,\mathrm{GeV}$. 
%as functions of temperature, obtained in physical point QCD. 
The vertical axis is in unit of ${\rm GeV}^3$ .
% Red circles are $u$ (or $d$) quark condensate and black triangles are that for $s$ quark. 
%Errors include both statistical and systematic errors. 
%Data at $T\simge247$ MeV ($N_t\le8$) may be contaminated by the $O\left( (aT)^2\right)=O\left(1/N_t^2 \right)$ lattice artifacts.
}
\label{fig_chiral_P}
\end{figure}

Our  preliminary results for the chiral condensates are given in Fig.~\ref{fig_chiral_P}. 
We find that, while the behavior of the strange quark condensate is similar to the case of heavy QCD shown in the left panel of Fig.~\ref{fig_chiral} and suggesting a crossover around 130-150 MeV, 
the light quark condensate shows a much sharper crossover/transition there.
Our preliminary results for their disconnected chiral condensates suggest that the peak locates at $T \simle 169$ MeV.
This is consistent with the results of the staggered quarks \cite{Borsanyi:2013bia,Bazavov:2014pvz}.

%----------------------------------------------------------------------------
\section{Summary}

Applying the method of Refs.~\cite{Suzuki:2013gza,Makino:2014taa,Hieda:2016lly} based on the gradient flow, we study thermodynamic properties of QCD with (2+1)-floavors of dynamical quarks.
As the first step, we studied QCD with heavy u,d quarks on a fine lattice with $a\simeq0.07$ fm.
Our EOS is consistent with that from the conventional $T$-integration method, suggesting that the lattice we have studied is sufficiently close to the continuum limit while the $O\left( (aT)^2\right)=O\left(1/N_t^2 \right)$ lattice artifacts are severe at $N_t\simle8$. 
We find that the gradient flow method is quite powerful in extracting physical properties even with Wilson-type quarks which violate the chiral symmetry explicitly and thus had not been easy to calculate chiral and topological properties with usual methods.

We are now extending the study to the physical point, though the lattice is slightly coarser with $a\simeq0.09$ fm.
Our preliminary results for the flow time dependence are similar to the heavy QCD case, but the linear windows are often narrower than the heavy case.
This requires a more careful study of systematic errors.
Our preliminary results at this single lattice spacing suggest $T_\textrm{pc} \simle 169$ MeV at the physical point.
To draw a more definite conclusion, we need data at lower temperatures and also at smaller lattice spacings.

\vspace{5mm}
%\noindent\textbf{Acknowledgments}

This work was in part supported by JSPS KAKENHI Grant Numbers
JP25800148, JP26287040, JP26400244, JP26400251, JP15K05041, JP16H03982, and JP17K05442.
This research used computational resources of HA-PACS and COMA provided by the Interdisciplinary Computational Science Program of Center for Computational Sciences at University of Tsukuba (No.\ 17a13), 
SR16000 and BG/Q by the Large Scale Simulation Program of High Energy Accelerator
Research Organization (KEK) (Nos.\ 13/14-21, 14/15-23, 15/16-T06, 15/16-T-07, 15/16-25, 16/17-05),
and Oakforest-PACS at JCAHPC through the HPCI System Research project (Project ID:hp170208).
This work was in part based on Lattice QCD common code Bridge++ \cite{bridge}.

%\clearpage
%\bibliography{lattice2017}

%%%%%%%%%%%%%%%%%%%%%%%%%%%%%%%%%%%%%%%%%%%%%%%%%%%%%%%%%%%%%%%%%%%%%%%%%%%%%
\end{document}